%% file: Arxiv_Main.tex
\documentclass[11pt, letterpaper]{article}
\usepackage[margin=1in]{geometry} 
\usepackage{chato-notes}
\usepackage{multirow}
\usepackage{subfig}
\usepackage{tikzpeople}
\usepackage{paralist}
\usepackage[square,numbers]{natbib}
\usepackage{authblk}
\usepackage{hyperref}
\usepackage[capitalise,noabbrev]{cleveref}

\AtBeginDocument{%
  \providecommand\BibTeX{{%
    \normalfont B\kern-0.5em{\scshape i\kern-0.25em b}\kern-0.8em\TeX}}}

\newcommand\blfootnote[1]{%
	\begingroup
	\renewcommand\thefootnote{}\footnote{#1}%
	\addtocounter{footnote}{-1}%
	\endgroup
}

\begin{document}

\title{Fair ranking: a critical review, challenges, and future directions\blfootnote{The views expressed in the paper should not be interpreted as reflecting those of authors' affiliated organizations. The authors would like to thank Francesco Fabbri, Jessie Finocchiaro, Faidra Monachou, Ignacio Rios, and Ana-Andreea Stoica for helpful comments. This project has been a part of the \href{https://www.md4sg.com/}{MD4SG} working group on Bias, Discrimination, and Fairness. This research was supported in part by European Research Council (ERC) Marie Sklodowska-Curie grant (agreement no. 860630), Spanish Ministerio de Ciencia, Innovación y Universidades (MCIU) and the Agencia Estatal de Investigación (AEI)-PID2019-111403GB-I00/AEI/10.13039/501100011033. G. K Patro acknowledges the support by TCS Research fellowship.}}

\author[1]{Gourab K Patro}
\author[2]{Lorenzo Porcaro}
\author[3]{Laura Mitchell}
\author[4]{Qiuyue	Zhang}
\author[5]{Meike Zehlike}
\author[6]{Nikhil Garg}
\affil[1]{IIT Kharagpur, India and L3S Research Center, Germany}
\affil[2]{Universitat Pompeu Fabra, Spain}
\affil[3]{Competition and Markets Authority, United Kingdom}
\affil[4]{Accenture Plc, United Kingdom}
\affil[5]{Zalando Research and Max Planck Institute for Software Systems, Germany}
\affil[6]{Cornell Tech, United States}

\maketitle
\begin{abstract}
Ranking, recommendation, and retrieval systems are widely used in online platforms and other societal systems, including e-commerce, media-streaming, admissions, gig platforms, and hiring.
 In the recent past, a large ``fair ranking'' research literature has been developed around making these systems fair to the individuals, providers, or content that are being ranked. Most of this literature defines fairness for a single instance of retrieval, or as a simple additive notion for multiple instances of retrievals over time.
This work provides a critical overview of this literature, detailing the often context-specific concerns that such an approach misses: the gap between high ranking placements and true provider utility, spillovers and compounding effects over time, induced strategic incentives, and the effect of statistical uncertainty. We then provide a path forward for a more holistic and impact-oriented fair ranking research agenda, including methodological lessons from other fields and the role of the broader stakeholder community in overcoming data bottlenecks and designing effective regulatory environments.  

\end{abstract}

\input{Arxiv_1_intro}
\input{Arxiv_2_ranking_fairness}
\input{Arxiv_3_pitfalls}
\input{Arxiv_4_towards_longterm}
\input{Arxiv_5_conclusion}

\bibliographystyle{abbrvnat}
\bibliography{Arxiv_Ref}

\input{Arxiv_6_appendix}

\end{document}

%% file: Arxiv_1_intro.tex
\section{Introduction}
Ranking systems are ubiquitous across both online marketplaces (e-commerce, gig-economy, multimedia) and other socio-technical systems (admissions or labor platforms), playing a role in which products are bought, who is hired, and what media is consumed.
In many of these systems, ranking algorithms form a core aspect of how a large search space is made manageable for \textit{consumers} (employers, buyers, admissions officers, etc). In turn, these algorithms are consequential to the \textit{providers} (sellers, workers, job seekers, content creators, media houses, etc.) who are being ranked.

Much of the initial work on such ranking, recommendation, or retrieval systems (RS\footnote{While we often use ``RS'' or ranking systems as shorthand, in this work we often mean ranking, recommendation, retrieval, and constrained allocation algorithmic systems more broadly -- systems that select (and potentially order) a subset of providers from a larger available set.}) focused on learning to maximize \textit{relevance}---often measured through proxies like clickthrough rate---, showing the most relevant items to the consumer, based solely on the consumer's objective \cite{liu2011learning,adomavicius2005toward}. However, like all machine learning techniques, such systems have been found to `unfairly' favor or discriminate against certain individuals or groups of individuals in various scenarios \cite{ekstrand2018all,BaezaYates2018,chen2020bias}.

Thus, as part of the burgeoning algorithmic fairness literature \cite{mehrabi2019survey,Mitchell2021}, there have recently been many works on fairness in ranking, recommendation, and constrained allocation more broadly \cite{burke2017multisided,zehlike2017fa, zehlike2022fair, geyik2019fairness, celis2018ranking, asudeh2019designing,singh2018fairness, biega2018equity, surer2018multistakeholder,guo2021stereotyping,cai2020fair}. For example, suppose that the platform is deciding how to rank 10 items on a product search result page, and each item has demographic characteristics (such as those of the seller). Then---in addition to considering each item's relevance---how should the platform rank the items, in a manner that is ``fair'' to the providers, either on an individual or group level? This question is often considered on an abstract level, independent of the specific ranking context; moreover, the literature primarily focuses on fairness of one instance of the ranking \cite{zehlike2017fa, zehlike2020reducing, zehlike2022fair, singh2018fairness}, or multiple independent instances of rankings with an additive objective across instances \cite{biega2018equity, suhr2019two}.

The goals of this paper are  to synthesize the current state of the fair ranking and recommendation field, and to lay the agenda for future work. In line with recent papers \cite{Jannach2020,Selbst2018} on both broader fairness and recommendations systems, our view is that the fair ranking literature risks being ineffective for problems faced in real-world ranking and recommendation settings, if it focuses too narrowly on an abstract, static ranking settings. To combat this trend, we identify several pitfalls that have been overlooked in the literature, and should be considered in context-specific ways: toward a broader, long-term view of the fairness implications of a particular ranking system.

Like much of the algorithmic fairness literature, fair ranking mechanisms typically are designed by abstracting away contextual specifics, under a ``reducibility'' assumption; i.e., many fair ranking problems of interest can be reduced to a standard problem of ranking, that is a set of items or individuals constrained to a chosen notion of fairness or optimized for a suitable fairness measure (or multiple instances of such ranking over time with simple additive extensions); however, as \citet{Selbst2018} elucidate, the abstractions necessary for such a reduction often ``render technical interventions ineffective, inaccurate, and sometimes dangerously misguided.''

\begin{figure}[t!]
	\center{
			\includegraphics[width=1\textwidth]{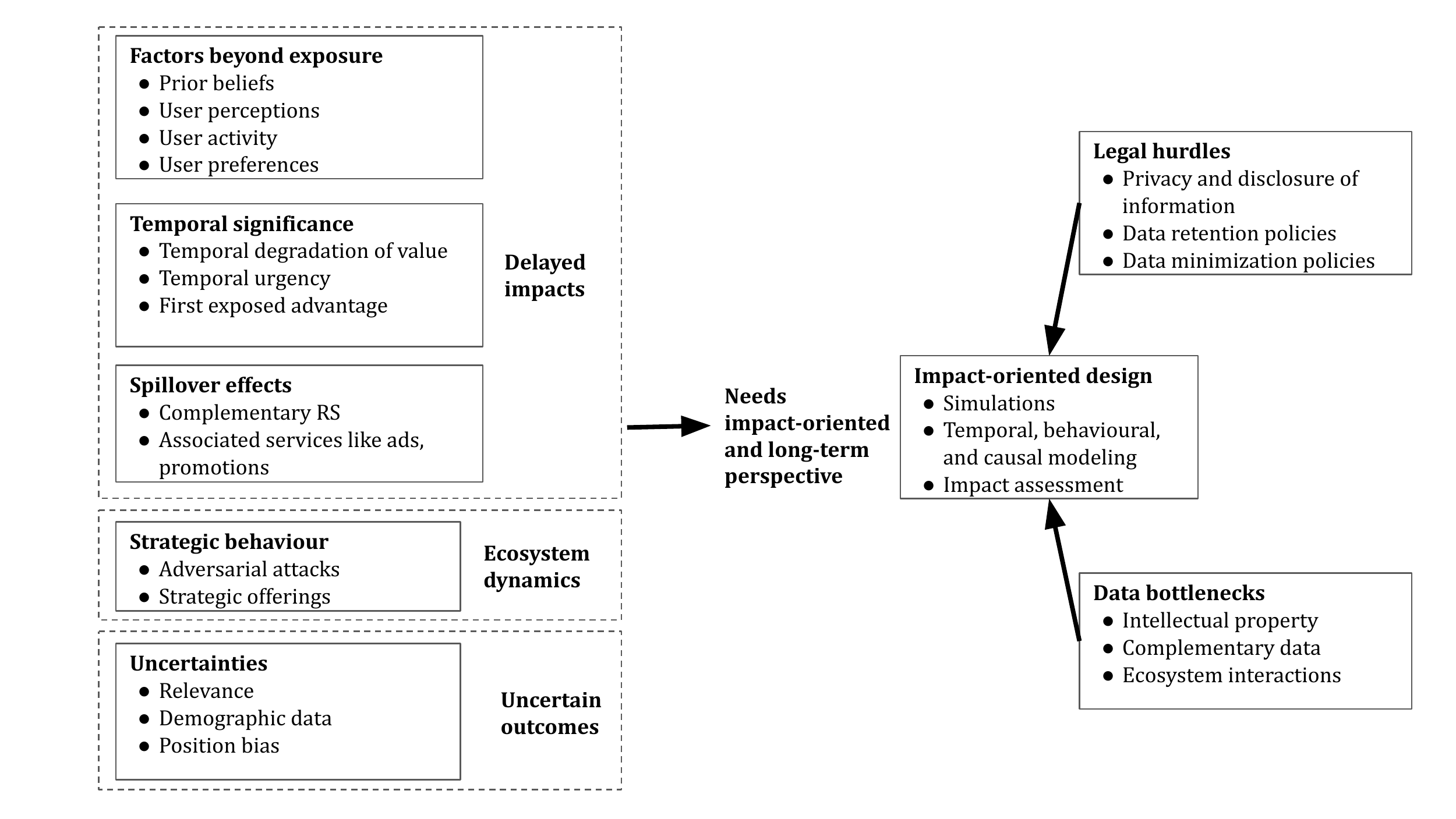}}
			\caption{This figure paints a big picture of the paper and succinctly summarizes our position on the field of fairness in retrieval systems, i.e., current fair RS mechanisms often fail to recognize several real-world nuances like delayed impacts, uncertainties in outcomes, ecosystem behaviour (discussed in \Cref{sec:pitfalls}); thus we must design fairness interventions in an impact-oriented approach with a holistic and long-term view of RS in mind. In \Cref{sec:long_term_fairness}, we discuss how algorithmic impact assessment can be helpful in this regard. More specifically in \Cref{subsec:simulations}, we overview various applied modeling techniques and simulation frameworks which in tandem can be used for impact-oriented studies of fairness in RS. Following this, in \Cref{subsec:data_bottlenecks,subsec:legal_bottlenecks} we briefly discuss various data bottlenecks and legal hurdles which might challenge the efforts towards a holistic view of RS fairness.}
			\label{fig:block_diagram}
\end{figure}

\textbf{Overview and Contributions}. In this work, we outline the many ways in which such a reduction often abstracts away many of the important aspects in the fair ranking context: the gap between position-based metrics and true provider utility, spillovers from one ranking to another across time and products, strategic incentives induced by the system, and the (differential) consequences of ranking noise. Studying fair ranking questions in such a reduced format and ignoring these issues might work in the ideal environment chosen during the problem reduction, but is likely insufficient to bring fairness in a real-world ranking system.
For example, a ranking algorithm that does not consider how relevance or consumer discrimination affects outcomes, or how early popularity leads to compounding rewards on many platforms, is unlikely to achieve its fairness desiderata; furthermore, ignoring strategic manipulation (such as Sybil attacks where a provider creates multiple copies of their profile or items) may lead to fairness mechanisms amplifying rather than mitigating inequities on the platform. We believe that these aspects must be tackled by the fair ranking literature, in order for this literature to positively affect practice. 

 We then overview methodological paths forward to incorporate these aspects into fair ranking research, as part of a broader long-term framework of algorithmic impact assessments---simulations, applied modeling, and data-driven approaches---along with their challenges. Finally, we conclude with a discussion on the broader regulatory, legal, and external audit landscape, necessary to translate the fair ranking literature into systems in practice.

\Cref{fig:block_diagram} summarizes our paper at a high level.

\textbf{Outline.} \Cref{sec:ranking_n_fairness} contains an overview of the fair RS literature. \Cref{sec:pitfalls} presents the aspects of ranking systems that we believe should be most covered by future fair RS work. \Cref{sec:long_term_fairness} contains the discussion of the paths forward within the broader data and regulatory landscape. 

%% file: Arxiv_2_ranking_fairness.tex
\section{Overview of Fair Ranking Literature}\label{sec:ranking_n_fairness}
Designing effective ranking, recommendation, or retrieval systems (RSs) requires tackling many of the same challenges as to build general machine learning algorithms---with additional challenges stemming from the characteristic that such systems make \textit{comparative} judgments across items; a high position in the ranking is a constrained resource.
RSs often employ machine learned models to estimate the {\it relevance} (or {\it probability of relevance}) of the items to any search or recommendation query \cite{liu2011learning,adomavicius2005toward}.
Historically, while user utility is the broader objective \cite{pu2011user}, the most popular guiding principle is the {\it Probability Ranking Principle} \cite{robertson1977probability}:
items are ranked in descending order of their probability to be relevant to the user, often estimated through click-through rates.
For a broad range of user utility metrics---such as mean average precision \cite{voorhees2000variations}, mean reciprocal rank \cite{voorhees1999trec}, and cumulative gain based metrics \cite{jarvelin2002cumulated,jarvelin2017ir}---this principle in turn maximizes the expected utility of users \cite{jarvelin2017ir}.

However, not only are more (estimated to be) relevant items typically ranked higher, but also users tend to click more on higher positioned items, even conditioned on relevance. 
Such a {\it position bias}  \cite{craswell2008experimental} means that expected attention (\textit{exposure}) from users decreases significantly while moving from the top rank to the bottom one; for example, users may evaluate items sequentially from the top rank, until they find a satisfactory one. It is thus important for producers to be ranked highly; a small difference in relevance estimation could result in a large difference in expected user attention (for example, see Appendix \Cref{tab:position_bias}). 
Depending on the ranking context, e.g., ranking products vs. ranking job candidates, high ranking positions directly translate to rewards, or at least increase their likelihood. (However, as we explain in the next section, the gap between exposure and true provider utility is an important one to understand.)

\textbf{Fairness in Rankings.} Due to the importance of rankings for providers,\footnote{Note that despite the recent explorations into multi-sided fairness in online platforms \cite{burke2017multisided,patro2020fairrec,suhr2019two}, we restrict our discussion to provider fairness which has been studied quite extensively.} and as part of the increased focus on machine learning injustices, there has been much recent interest in fairness and equity for providers rather than just ranking utility for consumers. 
There are numerous definitions, criteria and evaluation metrics to estimate a system's ability to be \textit{fair}  \cite{corbett2018measure,mehrabi2019survey,Mitchell2021,ekstrand2019fairness,distributive,castillo2019fairness,yao2017beyond}. 
Given heterogeneous settings, the complex environment in which retrieval systems are developed, and the multitude of stakeholders involved that may have differing moral goals \cite{finocchiaro2021bridging} and worldviews \cite{Friedler2021}, there is obviously no universal fairness definition; at a high level, however, many definitions can be classified into whether the objective is to treat similar individuals similarly (\textit{individual fairness}) \cite{dwork2012fairness}, or if different groups of individuals, defined by certain characteristics such as demographics, should be treated in a similar manner (\textit{group fairness}) \cite{speicher2018unified}.

In the following, we overview the concepts and works most relevant for our critiques and the agenda that we advocate.
Fairness notions from the domain of classification can---to a certain extent---be adopted to serve in ranking settings.
They typically only require additional consideration of the comparative nature of rankings and of how utility is modeled \cite{castillo2019fairness}. 
Compared to relevance-only ranking, adding fairness considerations often leads to the optimization of a multi-objective (or a constrained objective), where the usual utility (or relevance) objective comes along with a fairness constraint or objective focused on the providers \cite{Ribeiro2013,xiao2017fairness}.

One branch of the literature \cite{zehlike2017fa, zehlike2022fair, geyik2019fairness, celis2018ranking, asudeh2019designing} reasons about probability-based fairness in the top-$k$ ranking positions, which puts the focus onto group fairness.
These works commonly provide a minimum (and for some cases also maximum) number or proportion of items/individuals from a protected groups, that are to be distributed evenly across the ranking.
The methods do not usually allow later compensation, if the fairness constraints are not met at any of the top-$k$ positions (e.g., by putting more protected than non-protected items to lower positions).

Another set of works \cite{singh2018fairness, biega2018equity, surer2018multistakeholder,diaz2020evaluating, zehlike2020reducing} assign values (often referred to as {\it attention} or {\it exposure} scores) to each ranking position based on the expected user attention or click probability.
These works argue that the total exposure is a limited resource on any platform (due to position bias), and advocate for fair distribution of exposure to ensure fairness for the providers.
In contrast to the former line of work, using exposure as a metric to quantify provider utility has brought up not only group fairness notions~\cite{singh2018fairness,morik2020controlling}, but also definitions to enhance individual fairness~\cite{singh2018fairness,biega2018equity,bower2021individually}.
Further, in contrast to probability-based methods, these methods balance the \emph{total} exposure across individuals or groups, and thus they do allow compensations in lower positions.

Generally the problem definitions in these works center around a single instance of ranking, i.e., at a particular point in time we are given a set of items or individuals, their sensitive or protected attribute(s) (e.g., race and gender), and their relevance scores; the task is to create a ranking which follows some notion of fairness (like demographic parity or equal opportunity) for the items or individuals, while maximizing the user utility.
Some exceptions are \citet{biega2018equity}, \citet{suhr2019two} and \citet{surer2018multistakeholder}, that propose to deterministically ensure fairness through equity in amortized exposure, i.e., addition over time or over multiple instances of ranking.
In the next section, we argue that both these broad approaches (probability-based, and exposure-based) may be incomplete in many applications, due to their exclusive focus (either directly or indirectly) on ranking positions.

%% file: Arxiv_3_pitfalls.tex
\section{Pitfalls of existing fair ranking models}\label{sec:pitfalls}
In this section, we enumerate several crucial aspects of ranking and recommendation systems that substantially influence their fairness properties, but are ignored when considering an abstract fair ranking setting. The left hand side of \Cref{fig:block_diagram} summarizes this section. We begin in \Cref{subsec:beyond_exposure} by noting that exposure (or more generally, equating higher positions with higher utility) often does not translate to provider utility. \Cref{subsec:temporal_significance} discusses spillovers across rankings, either over time, across different rankings on the same user interface, or competition across platforms. \Cref{subsec:strategic_behavior} discusses strategic provider responses, and how they may counter-act (or worsen) the effects of a fair ranking mechanism. Finally, \Cref{subsec:uncertainty} illustrates how noise---either in demographic variables or in other aspects---may differentially affect providers within a fair ranking mechanism.

Note that these issues are also present in other aspects of ranking, and in algorithmic fairness literature more generally; in fact, we also discuss if and how such issues have been studied in related settings. However, we believe that the intersection of fairness and ranking challenges amplify these concerns; for example, the naturally comparative aspect of rankings worsens the effects of competitive behavior and differential uncertainties.

Finally, while these pitfalls may not be the only ones, we believe these are the major ones which may cause the failure of proposed fair ranking frameworks in delivering fair outcomes in several real-world scenarios.
In the next section (\cref{sec:long_term_fairness}), we elaborate on how to tackle these challenges.
\subsection{Provider Utility beyond Position-based Exposure}\label{subsec:beyond_exposure}
As discussed above, the fair ranking literature often uses \textit{exposure} as a proxy for provider utility\footnote{Note that, here we are talking about the utility gained by a provider as a result of getting ranked. Thus provider utility is not same as user utility.} \cite{ekstrand2019fairness, singh2018fairness, castillo2019fairness, zehlike2020reducing}. For example, well-known fair ranking mechanisms like {\it equity of attention} \cite{biega2018equity} and {\it fairness of exposure} \cite{singh2018fairness,zehlike2020reducing} emphasize fairly allocating exposure among providers. Such works often implicitly assume that exposure is measured solely through a provider's position in the ranking; i.e., each position is assigned a value, independent of context. While such ranking-position-based exposure is often a useful measure of provider utility, such a focus misses context-specific factors due to which higher exposure does not necessarily lead to increased user attention, or that increased user attention may not directly translate to provider utility, as measured through, e.g., sales or long-term satisfaction.

This measurement-construct gap---between exposure as a measurement and provider utility as the construct of interest---is not a challenge unique to fairness-related questions in ranking. 
For example, not distinguishing between varying levels of attention from users could affect the performance of algorithms designed to maximize sales, as it would affect the predictions of algorithms using exposure to calculate sales probabilities \cite{moe2004dynamic} or information diffusion on a social network \cite{bakshy2012role}.
However, this gap may be especially important to be considered in a research direction that often seeks algorithmic solutions to inequities stemming from multiple causes, including the actions of other platform participants; for example, much work has analyzed (statistical or taste-based) discrimination on online platforms in which, even conditional on exposure, one type of stakeholders are treated inequitably by other stakeholders (see, e.g., racial discrimination by employers \cite{edelman2017racial,monachou2019discrimination}). 
In such settings, fair-exposure based algorithms may not uniformly or even substantially improve outcomes (we give an example in Appendix \Cref{tab:fair_exposure_gone_wrong}); this was recently underscored by \citet{suhr2020does}, which found through a user survey that such algorithms' effectiveness substantially depends on context such as job description and candidate profiles.

Another especially relevant contextual factor beyond position is \textit{time}: in fast moving domains like media, items may only be relevant for a short period of time \cite{campos2014time,yuan2013time}. 
In such scenarios, the stakeholders (both users and providers) most benefit from immediate exposure. 
For example, recency is an important aspect of relevance in breaking news \cite{chakraborty2017optimizing}, job candidates should be shown before vacancies are filled, and restaurants get more orders if recommended during peak hours to nearby customers \cite{yuan2013time, Banerjee2020AnalyzingM}.

More broadly, one should consider \textit{which providers} are being exposed to \textit{which users} and \textit{when}, as the value of a ranking position depends substantially on such match relevance and participant characteristics. 
Fair ranking models focusing solely on position, and thus oblivious to such context, may not have the desired downstream effects and may fail to deliver on fairness. 
We illustrate this consequence in an example in Appendix \Cref{tab:temporal_significance}.

\subsection{Spillovers effects: compounding popularity, related items, and competition}\label{subsec:temporal_significance}

While the immediate effect of an item's position in the ranking (e.g., an immediate sale) may be first-order, there are often substantial \textit{spillover} effects or \textit{externalities}, which should be incorporated in fair RS models. 
Here, we discuss three of such effects: compounding popularity or first-exposed-advantage, spillovers across products and ranking types, and competition effects.

Perhaps the most important spillover is a \textit{compounding popularity} or \textit{first-exposed-advantage},\footnote{The phrase is used to indicate its similarity to the {\it first-mover-advantage} phenomenon \cite{kerin1992first}.} in which the exposure an item receives during its early stages can significantly affect its long-term popularity \cite{figueiredo2014dynamics}. 
For example, early feedback in terms of clicks, sales, etc. could improve an item's estimated relevance scores, raising its future rankings; there may further be a popularity bias or herding phenomenon in which users are more likely to select an item, if they observe that others have selected it before them \cite{steck2011item,abdollahpouri2017controlling,salganik2008leading}. 
Similarly, as reflected in re-targeting in advertising, user preferences may change with exposure to an item. 
Thus, past exposure plays a huge role in determining the long-term effects of future exposure; denial of early exposure could risk the viability of small providers \cite{mladenov2020optimizing}.
Though one may intuitively think that continuous re-balancing of exposure through fairness-enhancing methods may overcome (or at least reduce) this problem, the real-world-proof is still to be made and early evidence suggests otherwise (see \citet{suhr2020does}).

Second, ranking systems---such as product recommendations---are rarely deployed as stand-alone services. 
They are often accompanied by associated services such as sponsored advertisements \cite{hillard2010improving}, similar or complementary item recommendations on individual item pages on e-commerce, media-streaming platforms and other marketplaces \cite{pazzani2007content,lai2021understanding}, non-personalized trending items \cite{cremonesi2010performance,benhardus2013streaming,platt2015international}, and other quality endorsements like editor's choice \cite{holly2012play}. 
Due to the presence of these associated services, user attention reaching an item may spill over to other items \cite{liang2019spillover,raj2021friends}. 
For example, complementary items or items similar to an item may receive spillover exposure thereby resulting in increased exposure levels for such items, via `you may also be interested` or `items similar to' recommendations, potentially leading to undesirable inequalities even under a fair RS model; we give such an example in Appendix \Cref{tab:spillover_example}.

Finally, there are competition and cross-platform spillover effects \cite{krijestorac2020cross,farahat2016app}: users may reach an item, not through the recommendation engine on the platform, but, e.g., via a search engine \cite{jansen2006effectiveness}, product or price comparison sites \cite{jung2014online}, or other platforms like social media \cite{hoffman2010can,saravanakumar2012social}.
In these instances, the recommendation engine at the user entry-point, e.g., the search engine’s recommendation system, will have a downstream effect on the exposure of items on the end site where the items are listed. These spillover effects could be important to analyze when designing potential `entry-point' recommendation systems. Perhaps more importantly---since a platform does not have control over all the off-platform systems that may influence item exposure on its own platform---one should consider how such external sources affect both the goals and the behavior of a fair RS system. In this regard, the major questions which remain understudied and unanswered at large are: should a fair RS consider the inequities induced via external systems and seek to counteract through interventions or should it ignore these effects for the sake of free market competition?

Together, these spillover effects suggest that fairness in RS (especially in recommendations) should not be modeled in isolation from associated and external services, and must take into account how the recommendations may have downstream consequences over time and space for either the same provider or on other providers. We note that these spillover effects are analogous to the  \textit{Ripple Effect trap} as described by \citet{Selbst2018}, in which harmful effects often stem from the failure of understanding how the introduction of new technologies could alter behaviours and values in existing social systems.

\subsection{Strategic Behavior}\label{subsec:strategic_behavior}
Current fair ranking mechanisms often fail to consider that the providers themselves could be strategic players who might try to \emph{actively} maximize their utilities \cite{tennenholtz2019rethinking,bahar2015economic}.
Providers often have an incentive to suitably strategize their offerings, e.g., content creators on media platforms could leave their own area of expertise and try to copy other popular creators or follow the popular trends \cite{ben2018game,ben2020content}, sellers could perform data poisoning attacks (through fake reviews, views, etc.) on the RS to improve their ranking \cite{zhang2020practical}, influencers on social network sites could try to hijack popular trends \cite{goga2015doppelganger,chakraborty2019equality}.
Providers can even strategically exploit the deployed fair ranking mechanisms to extract more benefits \cite{frobe2020effect,diincentives}.
Not factoring in such strategic behavior could impact ranking and recommendation systems, and especially the performance of fair ranking mechanisms. 

In the following, we overview some examples of strategic behavior and their consequences.
As in the measurement-construct gap between exposure and producer utility, strategic behavior as a reaction to ranking models is not just a question of fairness. 
Numerous works suggest that relevance estimation models are highly vulnerable to various types of adversarial attacks:
\begin{inparaenum}
	\item \emph{shilling attacks}, in which a provider gets associated with a group of users who then add supportive reviews, feedbacks, clicks, etc. to manipulate rankings in favor of the provider \cite{lam2004shilling};
	\item \emph{data poisoning attacks}, where a provider strategically generates malicious data and feeds it into the system through a set of manipulated interactions \cite{li2016data,zhang2020practical}; or
	\item \emph{doppelganger bot attacks}, where a number of fake users or bots are created and then strategically placed in a social network to hijack news feed ranking systems in favor of the malicious party \cite{goga2015doppelganger,chakraborty2019equality,molavi2013iolaus}.
\end{inparaenum}

However, some strategic behavior may specifically exploit characteristics of fair ranking algorithms. For example, fair ranking mechanisms may incentivize \emph{content duplication attacks} \cite{frobe2020effect}.
Strategic providers can create duplicates or near-duplicates---possibly hard to automatically identify---of their existing offerings in a ranking system.
Since certain fair ranking mechanisms may try to ensure benefits for all listed items, providers with more copies of same items stand to gain more benefits \cite{frobe2020effect,diincentives}.
We give such an example in Appendix \Cref{tab:duplication_attack}.
Other `undesirable' strategic behavior includes the purposeful provision or withholding of information, which may help some participants maximize their ranking; For example, in admissions settings, test-optional admissions policies that aim to be fair to students without test access may inadvertently be susceptible to strategic behavior by students with access but low test scores~\cite{liutestoptional21}. 

Strategic behavior by providers need not always be malicious;
rather, it could also represent a sincere effort for improvement (e.g., effort to improve restaurant's quality \cite{luca2016reviews}) or just a change in content offering strategy (e.g., strategic selection of topics for future content production \cite{halvorson2012content,raifer2017information}).
However, such `legitimate' strategic behavior may nevertheless affect the efficacy of fair ranking mechanisms over time, as such behavior may affect the relative performance of marketplace participants.
For example, \citet{vonderau2019spotify} shows that providers on various content sharing platforms may partly or completely change their content production strategy to cater to the taste of a ranking algorithm (instead of the taste of users).
Studies by \citet{Chaney2018} and \citet{ben2020content} suggest that ranking mechanisms which are unaware of such behavior could cause homogenization of a platform's item-space and degrade user utility over time; such behavior could also risk the long-term viability and welfare of small-scale providers \cite{mladenov2020optimizing}. Theoretically, \citet{liu2021strategic} extend the strategic classification literature to the ranking setting, to show that such effort (and its differential cost) could have substantial equity implications on the ultimate ranking.
Fair ranking mechanisms which seek to equalize exposure affect such incentives, both for desirable and undesirable strategic behavior, and it is necessary to take them into account when designing fair ranking mechanisms for real world settings. 
Designing fairness mechanisms which can distinguish between such desirable and undesirable behavior may be further challenging (cf. \cite{liutestoptional21}).

Finally, we note that the above discussion---that of strategic behavior of individual providers---does not consider the setting in which the platform---a seemingly neutral player and deployer of a ranking algorithm---also plays the role of a competitive provider (through a subsidiary or partner).
Since such providers have access to private platform data and control over their algorithms, they may be able to deploy undetectable strategic manipulations (e.g., Amazon's private label of products on its marketplace \cite{dash2021umpire}) which the other providers are not able to match, leading to an unfair strategy playing field for providers. 
The design and auditing of ranking algorithms robust to such behavior is an important direction for future work.

\subsection{Consequences of Uncertainty}\label{subsec:uncertainty}
Fairness-aware ranking mechanisms proposed for exposure- and probability-based fairness often assume knowledge of true relevance of providers or items, demographic characteristics on which to remain fair and of the value of each position in the ranking. However, such scores are rarely available in real-world settings.
For example, machine-learned models or other statistical techniques used to estimate relevance scores are often uncertain about the relevance of items due to various reasons, for example, biased or noisy feedback, the initial unavailability of data \cite{morik2020controlling,yang2021maximizing}, and platform updates in dynamic settings \cite{patro2020incremental}. While such estimation noise (or bias) is important for all algorithmic ranking or recommendations challenges, it is especially important to consider for fair ranking algorithms, as we illustrate below.

Current fair ranking mechanisms assume the availability of the demographic data of individuals to be ranked.
Whilst such assumptions help algorithmic developments for fair ranking, the availability of demographic data can not be taken for granted.
Demographic data such as race and gender is often hard to obtain due to reasons like legal prohibitions or privacy concerns on their collection in various domains \cite{andrus2021we,bogen2020awareness}.
To overcome the data gap, platform designers often resort to data-driven inference of demographic information \cite{lahoti2020fairness}, which usually involves huge uncertainty and errors \cite{andrus2021we}; the use of such uncertain estimates of demographic data in fair ranking mechanisms can cause significant harm to vulnerable groups, and ultimately fail to ensure fairness \cite{ghosh2021fair}.
Moreover, in dynamic market settings where protected groups of providers or items are often set based on popularity levels, the protected group membership changes over time, thereby adding temporal variations in demographics along with the uncertainty issues \cite{ge2021towards}.
To tackle such variations, \citet{ge2021towards} propose to use constrained reinforcement learning algorithms which can dynamically adjust the recommendation policy to nevertheless maintain long-term fairness. However, incorporating such demographic uncertainty to broader fair ranking algorithms remains an open question.

Another crucial part of rankings systems is the estimation of position bias \cite{agarwal2019estimating,chandar2018estimating} which acts as a proxy measure for click-through probability and helps quantify the possible utilities of providers based on their ranks \cite{bar2009presentation}.
Fairness-aware ranking mechanisms need these position bias estimates to ensure fair randomized or amortized click-through utility (exposure) for the providers.
While these estimates are often assumed to be readily available in most of the recent fair ranking systems works \cite{singh2018fairness,biega2018equity,diaz2020evaluating}, it also has huge uncertainty attached since it heavily depends on the specifics of the user interface.
Dynamic and interactive user interfaces \cite{mesbah2012crawling} used on many platforms, usually go through automatic changes which affects the attention bias (position and vertical bias) based on changes in web-page layout \cite{oosterhuis2018ranking}.
Furthermore, factors like the presence of attractive summaries and highlighted evidences for relevance---often generated in automated manners---alongside ranking results also differentially affect click-through probabilities over time and across items \cite{yue2010beyond,joachims2017accurately}.
Finally, the presence of relevant images, their sizes, text fonts, and other design constraints also play a huge role \cite{liu2015influence, wang2016beyond,granka2004eye}.
Together, as also discussed in \citet{wang2018position} and \citet{sapiezynski2019quantifying}, inaccuracies in position bias estimation and corresponding consequences remain important challenges in fair RS.

Finally, we note that uncertainties, including the above, may be \textit{differential}, affecting some participants more than others, even within the same protected groups. 
Such differential informativeness, for example, might occur in ranking settings where the platform has more information on some participants (through longer histories, or other access differences) than others \cite{emelianov2020fair,garg2021standardized}. 
The result of such differential informativeness may cause downstream disparate impact, such as privileging longer-serving providers over newer and smaller ones.

Together, these sources and areas of uncertainty should be an important aspect of future work in fair ranking.

\vspace{1em}
\noindent{\textbf{Fair ranking desiderata. }}
What should a comprehensive and long-term view of fairness in RS and its dynamics be composed of?
First, the provider utility measure should look beyond mere exposure, and account for user beliefs, perceptions, preferences and effects over time (as discussed in \Cref{subsec:beyond_exposure}). 
Second, fair RS works should consider not just immediate impacts but also their spillovers, whether over time for the same item or spillover effects on other items (as discussed in \Cref{subsec:temporal_significance}).
Third, strategic behavior and systems incentives should also be modeled to anticipate manipulation concerns and their adverse effects (as discussed in \Cref{subsec:strategic_behavior}).
Finally, fair RS mechanisms should incorporate the (potentially differential) effects of estimation noise (as discussed in \Cref{subsec:uncertainty}).

Putting things together, this section illustrated various challenges and downstream effects of developing and deploying algorithms from the fair RS literature. As we discuss in the next section, overcoming these challenges requires both longer-term thinking---beyond the immediate effect of a ranking position---and moving beyond studying general RS settings to modeling and analyzing specific settings and their context-specific dynamics.

%% file: Arxiv_4_towards_longterm.tex
\section{Towards Impact-oriented Fairness in Ranking and Recommender Systems}\label{sec:long_term_fairness}
In order to avoid the pitfalls discussed in the last section and to design `truly' fair RS, one must understand and assess the full range and long-term effects of various RS mechanisms. In this regard, we apply recent lessons from and critiques of Algorithmic Impact Assessment (AIA), both within and beyond the FAccT community. Algorithmic Impact Assessment (AIA) can be described as a set of practices and measurements with the purpose of establishing the (direct or indirect) impacts of algorithmic systems, identifying the accountability of those causing harms, and designing effective solutions \cite{Metcalf2021,Reisman2018}. More specifically to ranking and recommendation systems,  \citet{Jannach2020} introduces a comprehensive collection of issues related to impact-oriented research in RS. There are two broad lessons from this literature, that we explain and apply to the design of fair RS, in a manner that involves integrated effort from different actors and a comprehensive view of their effects.

First, as discussed by \citet{Vecchione2021}, a key point when assessing or auditing algorithmic systems is to move \textit{beyond discrete moments of decision making}, i.e., to understand how those decision-points affect the long-run system evolution; this point is particularly true for fairness interventions in ranking and recommender systems, as discussed in \Cref{sec:pitfalls}. \citet{Jannach2020} also highlight the limitations and unsuitability of traditional research in RS, which focused solely on accurately predicting user ratings for items (``leaderboard chasing") or optimizing click-through rates. Thus, in \Cref{subsec:simulations}, we begin with a discussion of methodologies that can be used to study such long-run effects of fair RS mechanisms, that have been used to study other questions in RS fields -- mainly, simulation and applied modeling. We detail not only the useful frameworks but also potential limitations and challenges when studying fairness-specific questions.

Second, a key aspect of effective assessments is the participation of every suitable stakeholder, including systems developers, affected communities, external experts, and public agencies; otherwise, a danger is that the research community focuses on impacts most measurable by its preferred methods and ignores others \cite{Metcalf2021}. However, there are bottlenecks to such holistic work, especially for RS used in private or sensitive contexts. We discuss data availability challenges in \Cref{subsec:data_bottlenecks}. Then, in \Cref{subsec:legal_bottlenecks}, we overview various regulatory frameworks -- along with their limitations -- designed to govern RS or algorithmic systems in general, and hold them accountable. Researchers should contribute to tackling these challenges as well. 
\subsection{Simulation and Applied Modeling to Study Long-term Effects and Context-specific Dynamics}\label{subsec:simulations}
Many of the challenges discussed in \Cref{sec:pitfalls} are regarding impacts that do not appear in the short-term, immediately after a given ranking; for example, it may take time for strategic agents to respond to a ranking systems. These long-term impacts are difficult to capture without considering a specific context, or with solely relying on ``traditional'' metrics that assess instantaneous precision-fairness trade-offs.

Outside of fair ranking, the recommendations literature has investigated such long-term and indirect effects using \textit{simulation and applied modeling} methods, motivated for example by the observation that offline (and commonly, precision-driven) recommendation experiments are not always predictive of long-term simulation or online A/B testing outcomes \cite{gomez2015netflix,bodapati2008recommendation, krauth2020offline}. However, surprisingly, such an approach has been relatively rare in the fair rankings and recommendations literature; to spur such work, here we overview various simulation and modeling tools along that are advantageous in our context.

First, {\bf simulations} have already been used in the past to demonstrate long-term effects of recommender systems and search engines---although unrelated to fairness, in ways that static precision-based analyses can not. 
Examples are the demonstration of the {\it performance paradox} (users' higher reliance on recommendations may lead to lower RS performance accuracy and discovery) by \citet{Zhang2020}, the study of {\it homogenization} effects on RS users by \citet{Chaney2018}, a study on the emergence of {\it filter bubbles} \cite{Nguyen2014} in collaborative filtering recommendation systems and its impacts by \citet{Aridor2020}, the evaluation of reinforcement learning to rank for search engines by \citet{hu2018reinforcement}, and a study on {\it popularity bias} in search engines by \citet{fortunato2006topical}.
All relied on context-specific simulations of RS.
Many other works also leverage simulations \cite{Hazrati2020, Ferraro2020, patro2020incremental, Banerjee2020AnalyzingM, Bountouridis2019, DAmour2020, Yao2020, Mansoury2020, patro2020towards} to study various dynamics in recommender systems.
In summary, these works illustrate how simulation-based environments can help in {\it (i)} studying various hypothesized relationships between the usage of systems and individual and collective behavior and effects, {\it (ii)} detecting new forms of relationships, and {\it (iii)} replicating results obtained in empirical studies.

Given the usefulness of simulations, many simulation frameworks have been developed to study various fairness approaches for information retrieval systems; just to mention a few: MARS-Gym \cite{MARSGYM}, ML-fairness-gym \cite{DAmour2020}, Accordion \cite{McInerney2021}, RecLab \cite{krauth2020offline}, RecSim NG \cite{Mladenov2021}, SIREN \cite{Bountouridis2019}, T-RECS \cite{lucherini2021t}, RecoGym \cite{Rohde2018}, AESim \cite{gao2021imitate}, Virtual-Taobao \cite{shi2019virtual}.

Note however, that the simulated environments are created under certain assumptions on the interactions between the stakeholders and the system, which may not always hold in real-world. 
As emphasized by \citet{Friedler2021}, it is important to question how different value assumptions may be influential on the simulated environments, and which worldviews have been modeled while developing such frameworks. 
On a positive note, simulation frameworks can be designed to be flexible enough to give freedom in (de)selecting or changing the fundamental value assumptions in fair RS;
for example RecoGym \cite{Rohde2018} and MARS-Gym \cite{MARSGYM} provide freedom in setting various types of user behaviours and interactions with the system.
This flexibility allows impact and efficacy assessment under different ethical scenarios, and the study of fair RS mechanisms under various delayed effects and user biases (as discussed in \cref{subsec:beyond_exposure,subsec:temporal_significance}) -- we believe that leveraging such simulation frameworks is an important path forward to studying the various effects discussed above in a context-specific manner.

Second, various {\bf temporal, behavioural and causal models} have traditionally been used to formally define, understand and study complex dynamical systems in fields like social networks \cite{handcock2010modeling,hanneke2010discrete,farajtabar2017coevolve}, game theory and economics \cite{camerer2003behavioural,ariely2008predictably}, machine learning \cite{yao2021survey,guo2020survey}, and epidemiology \cite{grenfell2001travelling}. 
These models often rely on real-world observations of individual behaviour, extract broader insights, and then try to formally represent both individual and system dynamics through mathematical modeling.
While the simulation frameworks can function as technical tools to study RS dynamics, suitable temporal, behavioural and causal models can be integrated within the simulation to ensure that the eco-system parametrization, stakeholder behaviour and system pay-offs are representative of the real-world.
A good example: \citet{radinsky2013behavioral} try to improve search engine performance with the use of suitable behavioural and temporal models in their framework.
Similarly, simulation frameworks with suitable applied modeling can be used to design and evaluate fair RS mechanisms which can withstand strategic user behaviour and other temporal environment variations.
Causal models can be utilized to study the impact of fair RS \cite{sharma2015estimating, schnabel2016recommendations,wang2020causal} in presence or absence of uncertainties and various associated services. Applied modeling tools are further an effective way to study strategic concerns in ranking, along with their fairness implications \citep{liu2021strategic}.

Even though simulations along with applied modeling may not exactly mirror the real world effects of fair RS, they could give enough of a basis to highlight likely risks, which could then be taken into account while designing and optimizing fair RS mechanisms.
They also bring an opportunity to model the effects of proposed fairness interventions, so that their long-term and indirect effects can be better understood and compared.

However, these approaches would further benefit from availability of certain data and the resolution of related legal bottlenecks.
For example, studies on spillover effects can not proceed without the data on complementary and associated services.
These data and legal bottlenecks might have also contributed to the fact that there are very few works exploring this direction, and out of the limited works, some are limited to either theoretical analysis \cite{mladenov2020optimizing,ben2020content} or simulations with assumed parametrizations \cite{Zhang2020,ge2021towards,xue2019enhancing} in absence of complementary data.\footnote{Note that a few recent works look into long-term assessment of fair machine learning \cite{liu2018delayed,zhang2020long,DAmour2020}, which we overlook so as not to divert from the primary focus of our discussion.}
We discuss these bottlenecks in \cref{subsec:data_bottlenecks} and \cref{subsec:legal_bottlenecks}.
\input{Arxiv_4.2_data}
\input{Arxiv_4.3_legal}

%% file: Arxiv_4.2_data.tex
\subsection{\bf Data Bottlenecks}\label{subsec:data_bottlenecks}
A major challenge faced by researchers outside industry working on long-term comprehensive evaluations of fair RS is the unavailability of suitable data.

The traditional RS datasets \cite{harper2015movielens,mcfee2012million,bennett2007netflix,TREC_data} that often used in the literature were collected in times when goals like accuracy or click-through rates and so may not be a good fit for today's impact-oriented research \cite{Jannach2020}.
For example, a set of user-item ratings data such as the canonical MovieLens dataset \cite{harper2015movielens} may not capture how a user may value the item differently at different points in time or how a user's preferences evolve over time, or the user's or item's associated demographics.
Similarly, such data gives little insight into fake reviews or ratings \cite{luca2016fake,he2021market,li2016data,zhang2020practical}, or other strategic manipulations as discussed above. 
More broadly, such datasets do not include vital information such interface design changes that may have a behavioural impact on user choice (as discussed in \cref{subsec:uncertainty}), and associated services like complementary recommender systems or embedded advertisement blocks (as elaborated in \cref{subsec:beyond_exposure}) that work alongside the one being audited, the type and time of provider interactions and changes in their behaviour.
Such missing components of standard ranking and recommendation system datasets are a major bottleneck to studying the questions from \Cref{sec:pitfalls}.

On the other hand, the flourishing of the algorithmic fairness literature have contributed to the spread of several experimental datasets covering a wide range of scenarios such as school admission, credit score, house listings, news articles, and much more (see \cite{Zehlike2021survey, mehrabi2019survey} for a list of datasets used in fair ranking and ML research). Datasets such as \textit{COMPAS} or the \textit{German Credit} datasets, originally classification tasks, have been adapted to ranking settings.
A major issue related to the use of these datasets in fair ranking research is that they are often far from the contexts in which fair ranking algorithms would be used.
While potentially useful in the advancing the conceptual state-of-the-art in algorithmic fairness research, reliance on such datasets may raise significant concerns to the ecological validity of such research.
Therefore, a more detailed analysis on the use and characteristics of such datasets is a much needed work to address in future, similarly to what has been done in the context of Computer Vision research \cite{Miceli2021, Koch2021, Scheuerman2021}.

Here, we detail the characteristics that a RS dataset would need to be suitable for impact-oriented fairness analysis, in addition to the traditional indicators of user preference or experience (precision or click through rates). One recurring theme is that ranking and recommendation systems operate within a broader socio-technical environment (that they themselves shape), and existing datasets do not allow researchers to understand this broader environment and the underlying dynamics.\footnote{We note that while \textit{more} data is not always better (e.g., see the case of NLP models discussed by \citet{Bender2021})---we believe that a certain level of {\it completeness and richness of data} is required to perform more comprehensive and long-term impact analysis.} 
\begin{enumerate}[(1)]
\item Most easily, it would be useful to complement existing datasets with past data on the same platform, such as user-provider interactions and their behaviour; on RS's associated services and related rankings; on other contextual details such as user interface, page layout and design; and on past results from rankings, such as whether the user selected a custom sorting criteria like date or price instead of platform's default ranking criteria, whether the user was redirected to a product from an external or affiliate link, and whether the user's behaviour follows the platform's guidelines. Such complementary data would allow understand how the broader environment affects and is affected by a fair ranking algorithm. 
\item More broadly, a move from static datasets to temporal datasets -- with timestamps on ratings and displayed recommendations/ratings -- would allow finding temporal variations in RS and its stakeholders. It would further allow studying fairness beyond demographic characteristics, such as that related to new providers.  For example, as discussed in \Cref{subsec:temporal_significance}, higher ranked results can often lead to increased user attention and conversion rates \cite{craswell2008experimental}, i.e., results initially ranked higher could then have a greater chance of being ranked highly in subsequent rankings. Since such biased feedback could easily creep into temporal datasets, one must factor this in their RS impact analysis (e.g., an unbiased learning method by \citet{joachims2017unbiased} in presence of biased feedback). Studying such dynamics and their fairness implications in the real world requires observing such interactions.
\item Finally, as discussed in \Cref{subsec:uncertainty}, a key aspect of fairness in rankings is uncertainty, especially differential uncertainty. While some datasets may allow researchers to infer certain components of recommendation system uncertainty (such as by numbers of ratings for a provider), other uncertainties are hidden. External to such companies, it is unclear how to best reflect the correctness of provided user attributes (such as race and gender so as to avoid uncertainties in a platform's compliance to fairness), the genuineness of ratings and reviews (so as to account for manipulations in fair RS analysis) \cite{trustpilotrankeligibility,youtuberankeligibility}) when feedback is given, and other model uncertainties. While it may be difficult for companies to quantify their uncertainties when releasing datasets, one beneficial step would be to release more information on the origin of the data, i.e., dataset datasheets as described by \citet{Gebru2018}. 
\end{enumerate}

Unfortunately, as might be expected, there are several challenges to such comprehensive datasets. 

The most important challenges are from the legal domain, which might even affect researchers and developers within a company. For example, the data minimization principle in GDPR \cite{data_min_gdpr} could restrict platforms to collect sensitive information like gender or race, thereby indirectly closing the doors for the implementation of fairness interventions, and inferred attributes would contain huge uncertainty which may render fairness interventions useless (as discussed in \cref{subsec:uncertainty}).
In fact, a study by \citet{biega2020operationalizing} finds that the performance might not substantially decrease due to data minimization, but it might disparately impact different users.
Additional legal principles which might present challenges are other privacy regulations, data retention policies, intellectual property rights of platforms, etc.
We discuss these challenges in the next section.

Furthermore, while a comprehensive and long-term view on fair RS may be of huge societal need and expectation, the creation of suitable datasets and their availability to external researchers heavily rely on the interests of platform owners. Such external access, even if restricted in various ways, is an important aspect of regulation and auditing. 

We now turn to discussing such legal and regulatory concerns.

%% file: Arxiv_4.3_legal.tex
\subsection{\bf Legal Bottlenecks}\label{subsec:legal_bottlenecks}
In the previous section we discussed issues of missing data and the challenges to obtain necessary information due to platform interests and legal regulations on privacy. Regulations and other legal interventions by governments are helpful in some aspects of ensuring external audits, while hindering fair ranking and recommendation in other contexts. Legal provisions will vary across jurisdictions, causing different challenges in data access and algorithmic disclosure depending on the location of: the data requested, the users of platforms that implement RS’s, the individuals impacted by the rankings, and the researchers seeking access to RS information.  For example, data protection laws may potentially restrict access to data located in the EU, for non-EU based researchers or vice versa.

In this section we give an overview of legal hurdles that prevent researchers of fair RS from assessing the impact of their methods, along with information on specific laws and guidelines that can be used as a starting point for discussions to shape a more robust set of legal provisions for long term fair RS.

There are existing laws/guidance that could be applied to long term fairness in RS.  But the wording of some of these laws/guidance leaves them open to interpretation, such that a platform could reasonably argue that it is fulfilling its obligations under the guidance, without taking into account long term fairness in RS. The European Commission Ethics Guidelines for Trustworthy AI~\cite{EUEthicsGuidelines}  state that a system should be tested and validated to ensure it is working as intended throughout its entire life cycle, both during development and after deployment.  The guidelines list fairness as well as societal well-being as a requirement of trustworthy AI. 
However, if the word ``intended'' is interpreted narrowly, as point in time and in isolation from the dynamic and interconnected nature of recommendations, platforms could demonstrate that their systems are working as ``intended,'' considering both fairness and societal impact---even if in practice the platform may not be evaluating for long-term fairness or modelling various spillover effects.

In addition, the European Commission Guidelines on Ranking Transparency~ \cite{EURankingTransparency} reflect hesitancy that platforms have to be fully transparent on the details of their ranking; they recognise that providers are ``not required to disclose algorithms or any information that, with reasonable certainty, would result in the enabling of deception of consumers or consumer harm through the manipulation of search results.''
This privacy-transparency trade-off may cause the problem of missing data for algorithmic impact assessments to continue.

On the other hand, there is a push from regulators to make data from algorithmic systems available---if not to the general public, at least to independent third party auditors---to mitigate conflicts of interest when platforms audit their own systems.
In the US, the FTC’s Algorithmic Accountability Act \cite{FTCAlgorithmicAccountability} provides that if reasonably possible, impact assessments are to be performed in consultation with external third parties, including independent auditors and technology experts.
However, the EU harmonised rules for AI \cite{EUHarmonisedAIRules} acknowledge that given the early phase of the regulatory intervention and the fact the AI sector is very innovative, expertise for auditing is only now being accumulated.

In the absence of underlying data and full knowledge of the ranking algorithm, researchers could still adopt a forward looking approach of implementing simulations, based on what they do know about the ranking, to help predict the longer term effects of a ranking algorithm (as already explained in Section~\ref{subsec:simulations}).
It remains to be seen however, whether the advised disclosure of ``meaningful explanations'' of the main parameters of ranking algorithms---referred to in the European Commission Guidelines on Ranking Transparency \cite{EURankingTransparency}---provide enough information upon which to base an evaluation of the long term fairness of the RS.  There is also uncertainty over whether these meaningful explanations reduce sufficiently the impact of information asymmetry between users of the platform, and the platform itself, particularly where the platform both controls the RS, and includes its own items to be eligible in ranking results, alongside those of third party providers.  Further consideration also needs to be given to the timing of the release of the explanations when an RS method is updated, to give stakeholders sufficient opportunity to challenge reliance on these parameters, from a long term fairness perspective, pre-implementation of the RS update.

Applying laws to, or developing laws for, long term fairness scenarios in RS is in its infancy.  Those involved in shaping this legal framework should consider for long term fairness evaluation purposes: data access for different stakeholders, timings for this access, and level of detail that needs to be given; as well as providing actionable guidance on a platform’s responsibility for developing RS with long term fairness goals in mind.

%% file: Arxiv_5_conclusion.tex
\section{Conclusion}
In this paper we provided a critical overview of the current state of research on fairness in ranking, recommendations, and retrieval systems, and especially the aspects often abstracted away in existing research. 
Much of the existing research has focused on instant-based, static fairness definitions that are prone to oversimplifying real-world ranking systems and their environments.
Such a focus may do more harm than good and result in `fair-washing,' if those methods are deployed without continuous critical investigation on their outcomes.
Guidelines and methods to consider the effects of the entire ranking system through its life cycle, including effects from interactions with the outside world, are urgently needed.

We discussed various aspects beyond the actual ordering of items that affect rankings, such as spillover effects, temporal variations, and varying user characteristics ranging from their levels of activity. 
We further examined the effects of strategic behaviors and uncertainties in an RS.
These effects play an important role for the successful creation and assessment of fair rankings, and yet they are rarely considered in state-of-the-art fair ranking research.
Finally, we proposed next steps to overcome these research gaps.
As a promising first step we have identified simulations frameworks and applied-modeling methods, which can reflect the complexity of ranking systems and their environments.
However, in order to create meaningful impact analysis, concerns around datasets for fair ranking research, certain data bottlenecks and legal hurdles are yet to be resolved.

Our analysis concerning existing research gaps is of course by no means exhaustive, and many other issues of high complexity remain to be discussed.
In this paper, we focused on fair ranking methods that try to enhance fairness for a single side of stakeholders, mostly the individuals being ranked, or the providers of items that are ranked.
Research that is concerned with multi-stakeholder problems has recently started to emerge---finding, for example, that fairness objectives for providers and consumers in conflict to each other.

Similarly, we also did not explicitly discuss ranking platforms as two-sided markets, in which both sides may receive rankings for the other side.
While it is a promising direction with a vast corpus of economic research on the topic, it is important to understand that \begin{inparaenum}[(1)]
\item not all ranking platforms and their environments are two-sided in a literal sense: e.g., Amazon is a platform and a provider at the same time; and
\item depending on what is happening on the platform, different justice frameworks have to be applied: e.g., school choice, LinkedIn, and Amazon can all be seen as two-sided markets in a broader sense, but they need very different approaches when it comes to the question on what it means for them to be fair.
\end{inparaenum}
Depending on whether people or products are ranked, one might expect different user bias manifestations, as well as different requirements on data privacy and minimization policies. 
These differences have to be taken into account when designing fair ranking methods.

Finally, we note that, to the best of our knowledge, all known definitions of fairness in ranking are drawn from an understanding of fairness as distributive justice:
(limited) \textit{primary goods}---these are goods essential for a person's life, such as housing, access to job opportunities, health care, etc.---are to be distributed fairly across a set of individuals.
Fair ranking definitions of this kind may be a good fit for hiring or admissions, because we distribute a limited number of primary goods, namely jobs and education, among a set of individuals.
However, fairness definitions based on the distributive justice framework may not make sense in other scenarios.
For instance, e-commerce platforms may not qualify for properties of distributive justice, because they lack the aspect to distribute \emph{primary} goods: e-commerce settings, e.g., whether a single item is sold, may not qualify as immediately life-changing. 

Overall, we conclude that there is still a long way ahead of us; many more aspects from the ranking systems' universe have to be considered before we achieve substantive and robust algorithmic justice in rankings, recommendations, and retrieval systems.

%% file: Arxiv_6_appendix.tex
\newpage
\appendix
\section*{Appendix: Comprehensive Examples}
Here we give some toy examples relevant to our discussion in the paper.
\Cref{tab:position_bias} gives an example on how position bias in ranking could further widen the already existing inequalities. 
In \Cref{tab:fair_exposure_gone_wrong}, we give an example where the traditional fair ranking would fail to ensure equity in presence of user biases.
\Cref{tab:temporal_significance} gives an example where fair ranking mechanisms would fail in presence of temporal variations.
\Cref{tab:duplication_attack} and \Cref{tab:spillover_example} give examples on how duplication attacks and spillovers could cause the failure of fair ranking mechanisms.
\begin{table*}[h]
\small
\subfloat[An optimal ranking]{
\begin{tabular}{|c|c|c|c|c|}
	\hline
	{\bf Rank} & {\bf Expected} & {\bf Individual} & {\bf Relevance} & {\bf Group} \\
	& {\bf attention} & & & {\bf membership}\\\cline{1-5}
	$1$ & $0.5$ & \textcolor{blue}{A} & $0.92$ & \multirow{2}*{\textcolor{blue}{blue}} \\\cline{1-4}
	$2$ & $0.25$ & \textcolor{blue}{B} & $0.91$ & \\\cline{1-5}
	$3$ & $0.125$ & \textcolor{red}{C} & $0.90$ & \multirow{2}*{\textcolor{red}{red}}\\\cline{1-4}
	$4$ & $0.0625$ & \textcolor{red}{D} & $0.89$ & \\\cline{1-5}
\end{tabular}\label{tab:position_bias_ranking}}
\hfil
\subfloat[Group-level analysis]{
\begin{tabular}{|c|c|c|}
	\hline
	{\bf Group} & {\bf Mean} & {\bf Exposure}\\
	& {\bf relevance} & \\\cline{1-3}
	\textcolor{blue}{blue} & $0.915$ & $0.75$ \\\cline{1-3}
	\textcolor{red}{red} & $0.895$ & $0.1875$ \\\cline{1-3}
\end{tabular}\label{tab:position_bias_inequality}}
\caption{\textmd{Here we give an example (inspired by \citet{singh2018fairness}) on how position bias could further widen the existing inequalities. On a gig-economy platform there are four workers: A, B from the blue group, and C, D from the red group. For a certain employer, the platform wants to create a ranking of the workers. Let us assume that, in reality, all the workers are equally relevant to the employer. However, due to a pre-existing bias in historical training data, the relevance scores estimated by the platform's model are: $0.92$, $0.91$, $0.90$, $0.89$ for A, B, C, D respectively. Using the probability ranking principle \cite{robertson1977probability} we can optimize the user utility by ranking them in descending order of their relevance:
i.e., A$\succ$B$\succ$C$\succ$D as given in table (a). The second column of table (a) has the expected user attention for each rank (this follows from real world observations of position or rank bias indicating close to an exponential decrease of attention while moving from the top to bottom ranks \cite{craswell2008experimental}). Next we give a group-level analysis in table (b). The mean relevance scores of the blue group (A \& B) and the red group (C \& D) were $0.915$ and $0.895$ respectively which are not so different. On the other hand the exposure (sum of expected attention) of the blue and red groups---in the optimal ranking--- were $0.75$ and $0.1875$ respectively which are very different. We can clearly see that how the optimal ranking in presence of position bias, could significantly widen the gap in exposure even for a small difference in relevance estimation.}}\label{tab:position_bias}
\end{table*}

\begin{table*}[h]
\small
\subfloat[Expected attention]{
\begin{tabular}{|c|c|}
	\hline
	{\bf Rank} & {\bf Attention} \\\cline{1-2}
	$1$ & $0.6$ \\\cline{1-2}
	$2$ & $0.3$ \\\cline{1-2}
	$3$ & $0.1$ \\\cline{1-2}
\end{tabular}\label{tab:avg_attention}}
\hfil
\subfloat[Non-discriminatory employer]{
\begin{tabular}{|c|c|c|}
	\hline
	{\bf Rank} & {\bf Individual} & {\bf Group}\\\cline{1-3}
	$1$ & \textcolor{red}{A} & \textcolor{red}{red}\\\cline{1-3}
	$2$ & \textcolor{blue}{D} & \textcolor{blue}{blue} \\\cline{1-3}
	$3$ & \textcolor{blue}{E} & \textcolor{blue}{blue} \\\cline{1-3}
\end{tabular}\label{tab:ranking1}}
\hfil
\subfloat[Discriminatory employer]{
\begin{tabular}{|c|c|c|}
	\hline
	{\bf Rank} & {\bf Individual} & {\bf Group} \\\cline{1-3}
	$1$ & \textcolor{blue}{F} & \textcolor{blue}{blue} \\\cline{1-3}
	$2$ & \textcolor{red}{B} & \textcolor{red}{red} \\\cline{1-3}
	$3$ & \textcolor{red}{C} & \textcolor{red}{red} \\\cline{1-3}
\end{tabular}\label{tab:ranking2}}
\caption{\textmd{Here, we give a simple example of ranking in hiring or gig-economy platform setting where exposure, if used as a measure of producer utility, may fail to deliver desired fairness even after satisfying fairness of exposure. We have six workers (A, B, and C from the red group while D, E, and F from the blue group) on the platform. The platform's RS presents a ranked list (size $3$) of workers to the consumers i.e., the employers. In table (a), we give a sample distribution of expected attention from employer over the ranks (i.e., on average there are $0.6$, $0.3$, $0.1$ chances of an employer clicking on individual ranked $1$, $2$, $3$ respectively). Tables (b) and (c) show the rankings given to two different employers. Now the overall exposure of red group will be exposure$(A)+$ exposure$(B)+$ exposure$(C)= 0.6+0.3+0.1=1$. Similarly for blue group's exposure will be exposure$(D)+$ exposure$(E)+$ exposure$(F)= 0.3+0.1+0.6=1$. It is clear that this set of rankings follow the notions of fairness of exposure \cite{singh2018fairness} and equity of attention \cite{biega2018equity}. However, if we look more closely, one employer (in table (b)) is a non-discriminatory employer while the other one (in table (c)) is a discriminatory employer biased against the blue group. The second employer ignores the top ranked individual $F$ from blue group, and treats $B$, $C$ as if they are ranked at the first and second positions. Thus under these circumstances, the expected impact on the red group increases while that of the blue group decreases even though the rankings are fair in terms of exposure distribution.}}
\label{tab:fair_exposure_gone_wrong}
\end{table*}

\begin{table*}[h]
\small
\subfloat[Expected attention]{
\begin{tabular}{|c|c|}
	\hline
	{\bf Rank} & {\bf Attention} \\\cline{1-2}
	$1$ & $0.6$ \\\cline{1-2}
	$2$ & $0.4$ \\\cline{1-2}
\end{tabular}\label{tab:exp_exposure}}
\hfil
\subfloat[At Time $t$]{
\begin{tabular}{|c|c|c|c|}
	\hline
	{\bf Rank} & {\bf Item} & {\bf Exposure} & {\bf Overall}\\
	& & & {\bf interest}\\\cline{1-4}
	$1$ & $a$ & $0.6$ & \multirow{2}*{$1$} \\\cline{1-3}
	$2$ & $b$ & $0.4$ &  \\\cline{1-4}
\end{tabular}\label{tab:with_temporal_sig_1}}
\hfil
\subfloat[Time $t+1$ ($50\%$ reduction in overall interest)]{
\begin{tabular}{|c|c|c|c|}
	\hline
	{\bf Rank} & {\bf Item} & {\bf Exposure} & {\bf Overall} \\
	& & & {\bf interest}\\\cline{1-4}
	$1$ & $b$ & $0.6$ & \multirow{2}*{$0.5$} \\\cline{1-3}
	$2$ & $a$ & $0.4$ & \\\cline{1-4}
\end{tabular}\label{tab:with_temporal_sig_2}}
\caption{\textmd{Here we give an example on how temporal variations in the significance of rankings could fail the fair ranking mechanisms. Consider a scenario where two news agencies namely A and B regularly publish their articles on a news aggregator platform which then ranks the news articles while recommending to the readers (users). In table (a), we give a sample distribution of expected attention from readers over the ranks. At some time just before $t$, a big event happens, and both A and B quickly report on this through equally good articles $a$ and $b$, both published at time $t$. The table (b) and (c) show the rankings of articles on the platform at time $t$ and $t+1$. If we sum up the total exposure of each agency, we get exposure$(A)=0.6+0.4=1$ and exposure$(B)=0.4+0.6=1$. However, if we look more closely, the overall interest of readers on the breaking news at time $t$ is $1$ which decreases to $0.5$ at time $t+1$. This is because the readers who have already read the news on the particular event at $t$, will be less likely to read the same news again from a different agency at $t+1$. Thus, even though the exposure metric of the news agencies are the same in this case, they end up getting disparate impact due to the temporal degradation of user interest. A way to avoid such outcomes, would be to design and use suitable context-specific weighting mechanisms for rankings which can anticipate and account for such temporal variations.}}
\label{tab:temporal_significance}
\end{table*}

\begin{table*}[h]
\small
\subfloat[List of relevant items]{
\begin{tabular}{|c|c|c|}
	\hline
	{\bf Relevant} & {\bf Provider} & {\bf \% times}\\
	{\bf items}  & & {\bf recommended}\\\cline{1-3}
	$a_1$ & \multirow{2}*{$A$} & \multirow{2}*{$50\%$}\\\cline{1-1}
	$a_2$  & &\\\cline{1-3}
	$b_1$  & \multirow{2}*{$B$} & \multirow{2}*{$50\%$}\\\cline{1-1}
	$b_2$ & &\\\cline{1-3}
\end{tabular}\label{tab:normal_catalogue}}
\hfil
\subfloat[List with item duplication]{
\begin{tabular}{|c|c|c|}
	\hline
	{\bf Relevant} & {\bf Provider} & {\bf \% times}\\
	{\bf items}  & & {\bf recommended}\\\cline{1-3}
	$a_1$ & \multirow{3}*{$A$} & \multirow{2}*{$60\%$}\\\cline{1-1}
	$a_1$\_copy &  & \\\cline{1-1}
	$a_2$  & &\\\cline{1-3}
	$b_1$  & \multirow{2}*{$B$} & \multirow{2}*{$40\%$}\\\cline{1-1}
	$b_2$ & &\\\cline{1-3}
\end{tabular}\label{tab:manipulated_catalogue}}
\caption{\textmd{An example of duplication attack (inspired by \citet{diincentives} and the Sybil attacks in networks \cite{goga2015doppelganger}): Here, the table (a) has a list of relevant items for certain information need. The list contains two items each from providers $A$ and $B$. Let us consider a recommender system which recommends exactly one item every time. In such a recommendation setting, the fairness notions which advocate for fair allocation of exposure, visibility, or impact \cite{singh2018fairness,biega2018equity,surer2018multistakeholder}, would try to allocate $25\%$ to each item, i.e., each item is recommended $25\%$ of the time; thus each provider gets $50\%$ of the exposure or visibility. Now, if provider $A$ tries to manipulate by introducing a copy of its own item $a_1$ as a new item $a_1$\_copy (as shown in table (b)) potentially undetectable by the platform, then it is highly likely that the machine learned relevance scoring model would assign same or similar relevance to the copied item. In this scenario, due to the fairness notion, provider $A$ potentially increases her share of exposure to $60\%$ while reducing it to $40\%$ for provider $B$. Allocation-based fair ranking methods can create incentives for providers to do such strategic manipulations. Possible ways to dis-incentivise such duplication would be to actively include those item features in the relevance scoring model which are particularly harder to duplicate (e.g., \#views on YouTube videos, \#reviews on Amazon).}}
\label{tab:duplication_attack}
\end{table*}

\begin{table*}[h]
\small
\subfloat[Items and recommendations]{
\begin{tabular}{|c|c|c|}
	\hline
	{\bf Relevant} &  {\bf \% times}\\
	{\bf items}  & {\bf recommended}\\\cline{1-2}
	$a$ & $20\%$\\\cline{1-2}
	$b$ & $20\%$\\\cline{1-2}
	$c$ & $20\%$\\\cline{1-2}
	$d$ & $20\%$\\\cline{1-2}
	$e$ &  $20\%$\\\cline{1-2}
\end{tabular}\label{tab:item_catalogue}}
\hfil
\subfloat[Similar items]{
\begin{tabular}{|c|c|c|}
	\hline
	{\bf Item} & {\bf Similar items}\\
	{\bf page}  & {\bf recommended}\\\cline{1-2}
	$a$ & $b,c$\\\cline{1-2}
	$b$ & $c,d$\\\cline{1-2}
	$c$ & $a,b$\\\cline{1-2}
	$d$ & $b,c$\\\cline{1-2}
	$e$ & $b,c$\\\cline{1-2}
\end{tabular}\label{tab:similar_items_list}}
\hfil
\subfloat[Resultant exposure distribution]{
\begin{tabular}{|c|c|c|}
	\hline
	{\bf Item} & {\bf Resultant exposure}\\
	& {\bf (with $20\%$ spillover)}\\\cline{1-2}
	$a$ & $20-4+1\times 2=18\%$\\\cline{1-2}
	$b$ & $20-4+4\times 2=24\%$\\\cline{1-2}
	$c$ & $20-4+4\times 2=24\%$\\\cline{1-2}
	$d$ & $20-4+1\times 2=18\%$\\\cline{1-2}
	$e$ & $20-4+0\times 2=16\%$\\\cline{1-2}
\end{tabular}\label{tab:resultant_exposure}}
\caption{\textmd{An example on exposure spillover: Here we consider an e-commerce setting where there are five items relevant to a certain type of users. Following fairness of exposure \cite{singh2018fairness} or equity of attention \cite{biega2018equity}, each of the five items gets recommended same number of times as shown in table (a). Apart from this regular recommendation, e-commerce platforms often have similar or complementary item recommendations \cite{amazon_reco,sharma2015estimating} towards the bottom of individual item pages. Table (b) shows the similar items shown on each individual item's page. Assuming that there is 20\% spillover, i.e., 20\% of the user crowd coming to any item page moves to the similar items shown on the page, the resultant expected exposure of the items after one step of user spillovers is given in table (c). It can be clearly seen that even though the regular recommender system ensures fairness (as in table (a)), the resultant effects may not be fair due to spillover effects.}}
\label{tab:spillover_example}
\end{table*}